\documentclass[journal]{rmaa}
\usepackage{hyperref}

\usepackage{paralist}
 
\usepackage{psfrag,color,ulem}

\usepackage[latin1]{inputenc}

\usepackage{textcomp}

\usepackage{xspace} 
\newcommand{\vs}{V$_{s}$\/}
\newcommand{\cmc}{\,cm$^{-3}$\xspace}
\newcommand{\muG}{\,$\mu$G\xspace}

\newcommand{\kms}{\,km s$^{-1}$\xspace}

\newcommand{\forba}[3]{[\ion{#1}{#2}]\ #3\AA\xspace}
\newcommand{\sforba}[3]{\ion{#1}{#2}]\ #3\AA\xspace}

\newcommand{\forbm}[3]{[\ion{#1}{#2}]\ #3$\mu$m\xspace}

\newcommand{\OII}{\,[\ion{O}{ii}]\xspace}

\newcommand{\OIII}{\,[\ion{O}{iii}]\xspace}

\newcommand{\NII}{\,[\ion{N}{ii}]\xspace}

\newcommand{\SII}{\,[\ion{S}{ii}]\xspace}

\newcommand{\Ha}{\,H$\alpha$\xspace}
\newcommand{\Hb}{\,H$\beta$\xspace}

\newcommand{\VersionTwo}[1]{\textcolor{black}{\textbf{#1}}}

\newcommand{\M}{\textsc{mappings} }

\newcommand{\MIII}{\textsc{mappings\,iii} }

\newcommand{\MV}{\textsc{mappings\,v} }

\usepackage{listings}
\DeclareFixedFont{\ttb}{T1}{txtt}{bx}{n}{10} 
\DeclareFixedFont{\ttm}{T1}{txtt}{m}{n}{10}  

\usepackage{color}
\definecolor{deepblue}{rgb}{0,0,0.5}
\definecolor{deepred}{rgb}{0.6,0,0}
\definecolor{deepgreen}{rgb}{0,0.5,0}
\definecolor{darkgreen}{rgb}{0,0.6,0}

\newcommand\sqlstyle{\lstset{
language=SQL,
basicstyle=\ttm,
otherkeywords={self},             
keywordstyle=\ttb\color{deepblue},
emph={MyClass,__init__},          
emphstyle=\ttb\color{deepred},    
stringstyle=\color{deepgreen},
frame=tb,                         
showstringspaces=false            %
}}

\lstnewenvironment{sql}[1][]
{
\sqlstyle
\lstset{#1}
}
{}

\newcommand\pythoninline[1]{{\pythonstyle\lstinline!#1!}}

\newcommand{\sqlTab}[1]{{\fontfamily{txtt}\selectfont #1}}

\usepackage{threeparttable}

\title{Extensive Online Shock Model Database}

\author{
  A. Alarie\altaffilmark{1} \&
  C. Morisset\altaffilmark{1}}

\altaffiltext{1}{Instituto de Astronom\'{\i}a, Universidad Nacional Aut\'onoma de M\'exico, M\'exico}

\shortauthor{Alarie \& Morisset}
\shorttitle{Shock models database}

\fulladdresses{
\item A. Alarie: Instituto de Astronom\'{\i}a, Universidad Nacional Aut\'onoma de M\'exico, Apdo. Postal 70264, C.P. 04510, M\'exico, CDMX, M\'exico (alexandre.alarie@gmail.com).
\item C. Morisset: Instituto de Astronom\'{\i}a, Universidad Nacional Aut\'onoma de M\'exico, Apartado postal 106, C.P. 22800 Ensenada, Baja California, M\'exico (chris.morisset@gmail.com).}

\listofauthors{A. Alarie \& C. Morisset}
\indexauthor{Alarie, A.}
\indexauthor{Morisset, C.}

\abstract{We present a new database of fully radiative shock models calculated with the shock and photoionization code \textsc{mappings v}. The database architecture is built to contain diverse shock grids comprising of multiple shock parameters. It can be easily accessible through the MySQL protocol. Intensities of spectral lines from infrared to X-rays are stored along with other useful outputs such as the ionic fractions/temperature, integrated densities, etc. A web page was created in other to explore interactively the database as it evolves with time. Examples of its usage is given using the Python language.}

\resumen{}

\addkeyword{astronomical databases}
\addkeyword{H~II regions}
\addkeyword{ISM: abundances}
\addkeyword{Galaxy: abundances}

\hypersetup{draft}
\begin{document}
\maketitle

\section{Introduction}
\label{sec:intro}

The internet revolution has changed in a powerful and effective  way how research is being conducted today. Far away are the days when a researcher had to spend an incalculable amount of time in libraries searching and deciphering through an ever growing and complex scientific literature. In this day and age, researchers' primary reflex is to use the internet to search and gather essential information to the advancement of their research. Astrophysics has been particularly at the forefront in adopting this technology and applying it to very diverse goals.

Services such as the ADS\footnote{\url{http://http://adsabs.harvard.edu/}} (SAO/NASA Astrophysics Data System Abstract Service \citep{McKiernan+2001}) now plays an indispensable role in providing easy access to millions of abstracts and to the associated papers. 

Other web services have been created to facilitate the search of astrophysical data. A prominent example is the conglomerate of CDS\footnote{\url{http://cds.u-strasbg.fr/}} (Strasbourg astronomical Data Center) services like VizieR\footnote{\url{http://vizier.u-strasbg.fr/}} which became available in 1996 and was later described in \citet{2000A&AS..143...23O}. It provides access to the most complete online library of published astronomical catalogues and data tables organized in a self-documented database system.

Another service from CDS is the ALADIN\footnote{\url{https://aladin.u-strasbg.fr/aladin.gml}} interactive sky atlas \citep{Bonnarel+2000,Boch+2014} that allows simultaneous access to digitized images of the sky, astronomical catalogs, and databases. It is mainly used to facilitate  direct comparison of observational data at any wavelength with existing reference catalogs of astronomical objects. 

While such services have proven to be valuable, other areas of astrophysics can also benefited from those innovations. This is particularly the case of computational models such as photoionization and shock models computed using various spectral synthesis codes. Most of those models available on the 'market' can be found in the form of tables that are scattered around in the published literature. The most recent model grids are available as compressed files in multiple websites that use quite different data-formats. 
A centralized database that is readily accessible and user friendly would no doubt be beneficial to the community. To this end the Mexican Million Models database (3MdB\footnote{\url{https://sites.google.com/site/mexicanmillionmodels/}}) was created \citep{Morisset+2015}. It is designed to store and to distribute photoionization models computed with the code \textsc{cloudy} \citep{Ferland+2017} using the MySQL database management system. This service offers to the community an easy access to millions of online models by means of the \textsc{SQL} language.

This paper deals with the addition of shock models calculated with the code \textsc{mappings} \citep{Sutherland+17}. This new database which includes shock models is called "3MdBs" (i.e., 3MdB-shocks). 
The structure of the 3MdBs database differs from the original 3MdB (photoionization models) but the logic behind its usage has remained the same. Both databases are available at the same address and both can be used simultaneously using the appropriate \textsc{SQL} queries. 3MdBs comes with a website\footnote{\url{http://3mdb.astro.unam.mx}} which allows one to explore the grid available in the database interactively using a simple web browser. The website furthermore contains tutorials allowing potential users to obtain the necessary information required to interact with the database.

This paper is structured as follows. In Section\,\ref{sec:modelCode} we briefly introduce the modelling code \textsc{mappings\/}. Section \ref{sec:the_database} explains the database structure. The grids of models available at the time of publication are presented in section \ref{sec:grids_of_models} followed by a discussion about specials grids in section \ref{sec:particular_grids}. 

\section{The modeling code}
\label{sec:modelCode}

All the models referred to in this paper have been calculated using the shock and photoionization code \textsc{mappings\,v}{\footnote{\MV \VersionTwo{is freely accessible from \url{https://mappings.anu.edu.au/code/}}}}, version 5.1.13 \citep{Sutherland+17, Sutherland+18}. The latest improvements made in \MV are detailed in \citet{Sutherland+17}. 

\subsection{Preionization}
\label{sec:preionization}

Preionization conditions of the gas entering the shock front are an essential parameter of the shock calculations since it greatly influences the ionization structure of the shocked gas downstream and therefore all the line emissivities and their spatially integrated intensities \citep{Dopita+1995,Allen+08}. Manually setting preionization would be arbitrary and far from optimal, in particular in models related to specific astrophysical situations. 

\citet{Allen+08} for instance used an iterative process to determine preionization by first integrating the UV radiation propagating upstream that is produced by the shocked gas UV emission downstream, and second, by using the resulting UV energy distribution to calculate a photoionization model for the preshock gas as it travels towards the shock front. In the case of  high velocity shocks, it is a reasonable to assume that the precursor is photoionised and near equilibrium conditions as determined from the ionizing radiation field generated by the shocked gas downstream.

Since there is a strong feedback between preionization conditions and the ionization of the shocked gas downstream and its UV emission, the adopted methodology consisted in repeating the shock calculations using the preionization conditions inferred from the previous shock iteration. By repeating this iterative procedure up to at least four times, one finds that both the temperature and ionization state of the precursor converge towards a stable value, hence so do as well the line emission intensities of the cooling shock.

While this method is valid for shock velocities \vs\ in excess of $\sim$200\,\kms, it fails for slower velocities since there is insufficient time for the preshock gas to achieve equilibrium before it is shocked. In such case, it is essential to take into consideration the time dependent aspects of the problem. \MV\ now addresses the preionization problem in a fully consistent manner. For the first time, the new code treats the preshock ionization and thermal structure iteratively by solving, in a fully time-dependent manner, the photoionization of the preshock gas, its recombination, photoelectric heating and line cooling as it approaches the shock front. A detailed study of preionization in radiative shocks was presented in \citet{Sutherland+17}. Therefore, the models in the database make use of this new treatment in which the preshock temperature and ionization state of the gas are iteratively calculated after each shock calculation.

\subsection{Precursor gas}

For shocks with velocity less than 100 \kms, we did not compute the emission of the precursor since shocks below this velocity are unable to produce any appreciable ionizing emission. For shocks with velocity greater than 100 \kms, the photoionized precursors were computed separately. These were evaluated after the ionizing radiation field generated by the shock has been computed. All precursors were calculated subsequently to each iteration using the option 'P6' in \MV (photoionization model), the ionizing radiation emanating from the shock being the only source of ionization in the current grid. We essentially used the exact same method employed by \citet{Allen+08} to compute the precursor emission spectrum.

\section{The new database}
\label{sec:the_database}

All models presented in this papers are stored into an SQL\footnote{Structured Query Language} database freely available online. This method of distribution presents three main advantages. First, it allows anyone to have access instantaneously to thousands of models without the hassle of installing and managing the database on one's own workstation. Second, any new database updates or additions become instantly available to the community, a feature which can be very useful when it involves a worldwide collaboration (which is not available when using the VizieR database system). Third, the database can be accessed and  handled using anyone's preferred programming language as long as it includes a SQL client library. \\ 

In order to make use of this database and exploit fully its capabilities, the user needs to be familiar with the SQL database language, a domain-specific language designed for managing data stored in a relational database management system. This work makes use of MariaDB\textsuperscript{\tiny\textregistered}, a fast, scalable and robust open source database server. In the following section, we shall explain the database structure and the variables the latter manages. This will be followed by a book case example of how the database is best used. 

\subsection{Database structure}
\label{sec:database}

\begin{figure}[!t]
  \includegraphics[width=\columnwidth]{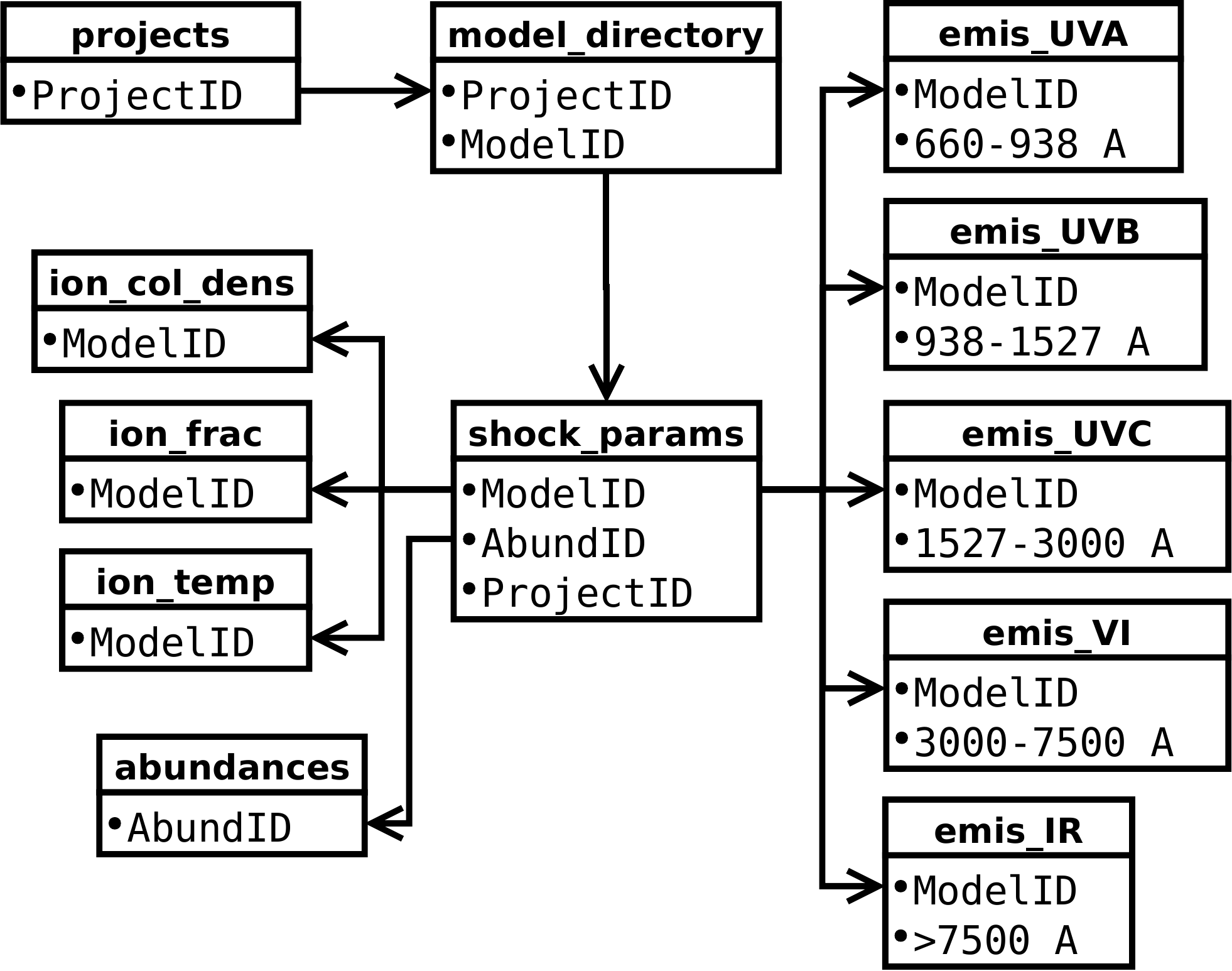}
  \caption{Interrelation between the various database Tables.}
  \label{fig:diagramDatabaseStructure}
\end{figure}

As in any database design, the data in distributed across several tables with a multicolumn setup. The whole database consists of 12 tables, each having a specific purpose which we shall discuss in details below. An overall view of the database's structure is displayed in Fig. {\ref{fig:diagramDatabaseStructure}}, which illustrates the different relations between the tables. The architecture adopts a modular design in the form of multiple tables, all with the aim of satisfying three criteria: efficiency, simplicity and expandability. By design, an SQL database can only contain a given number of name columns depending of its configuration. Too much columns in a table can greatly degrade its performance and even make it unstable. It is also very rare that one needs the information contained in all tables through a single database request. Most of the time, only a fraction of the table columns are needed at a given time. For this reason, the data are scattered into different tables, each having a specific purpose while limiting data duplication and thus optimizing the database size. 

All the shock parameters are to be found in the table named \sqlTab{shock\_params} (Table \ref{tab:shockParamsTable}). This table includes various columns associated with the shock parameters described in Section\,\ref{sec:AllenGrid}, mainly the shock velocity (\sqlTab{shck\_vel}), the preshock density (\sqlTab{preshck\_dens}) and the transverse magnetic field (\sqlTab{mag\_fld}). The abundance sets used for a specific model is referenced using an abundance identification code (\sqlTab{AbundID}) which is linked to the table \sqlTab{abundances} that contains a list of all abundance sets available in the database at the time it is consulted. 

A total of 4132 emission lines from ultraviolet to infrared can be extracted from the database. They are distributed among 5 tables, each covering specific wavelength intervals : 660-938 \AA\xspace (\sqlTab{emis\_UVA}), 939-1527 \AA\xspace (\sqlTab{emis\_UVB}), 1528-2999 \AA\xspace (\sqlTab{emis\_UVC}), 3000-7499 \AA\xspace (\sqlTab{emis\_VI}) and 7500\AA\xspace-609$\mu$m (\sqlTab{emis\_IR}). The complete emission lines list can be consulted via the web-interface. 

For each emission line of every model, the line intensities corresponding to the three different region types (namely shock, precursor and shock+precursor) are available, using the filter 
\sqlTab{WHERE emisVI.modeltype="shock"} or 
\sqlTab{"precursor"} or \sqlTab{"shock\_plus\_precursor"} 
respectively.

The mean temperatures, weighted by the ionic fraction of the specie involved, are given for each model in table \sqlTab{temp\_frac}, with the average ionic fraction listed in the column labelled \sqlTab{ion\_frac} and the ionic column densities in the column \sqlTab{ion\_column\_frac}.

There are two more tables in the database that are usually not expected to be required. Their names are \sqlTab{projects} and \sqlTab{models\_directory}. They ought to be used only when one needs to reevaluate a model or when the latter contains information required by the web-interface. 

\subsection{Website and user credentials}

In section \ref{sec:grids_of_models}, we will describe the models grids available at the time of publication. With time, other grids will be calculated and added in the database. In order to publicized these grids to the community, with have created a website allowing to visualized the different grids available at the time of consultation. The website can be reached via \url{http://3mdb.astro.unam.mx/}.

The user credentials can be obtained via the website alongside with the IP adress and port needed to connect to the database.

The website is connected directly to the MariaDB\footnote{\url{https://mariadb.org/}} database. This means that it is adaptive and any changes made to the database will automatically appears into the different sections. New grids of models can be added to the database, which then become automatically visible and  accessible to the general public. Since each grid is built using distinct range of shock parameters, and the spacing between successive values is at times non uniform from grid to grid, the website provides a parameter explorer that allows any user to explore the database and, in an intuitive manner, compose relatively complex SQL queries, such as the one given as example in Table \ref{tab:sqlquery}. The website also includes interactive tutorials that aim at teaching any user about how to connect and interact with the database using the Python programming language.

\begin{table*}
\caption{Example of SQL command lines used to generate Fig.~2 diagram.}
\begin{sql}
SELECT shock_params.shck_vel AS shck_vel,
       shock_params.mag_fld AS mag_fld,
       log10(emis_VI.NII_6583/emis_VI.HI_6563) AS NII_Hb,
       log10(emis_VI.OIII_5007/emis_VI.HI_4861) AS OIII_Hb
FROM shock_params
INNER JOIN emis_VI ON emis_VI.ModelID=shock_params.ModelID
INNER JOIN abundances ON abundances.AbundID=shock_params.AbundID
WHERE emis_VI.model_type='shock'
AND shock_params.ref='Allen08'
AND abundances.name='Allen2008_Solar' 
AND shock_params.shck_vel<=200
AND shock_params.preshck_dens=1
ORDER BY shck_vel, mag_fld;
\end{sql}
\label{tab:sqlquery}
\end{table*}

\section{The Models Grids in the Database}
\label{sec:grids_of_models}

At the time of publication, 3 main grids are available in the database :

\begin{enumerate}
  \item An exact replica of \citet{Allen+08} grids (section \ref{sec:AllenGrid}). 
  \item An extension of the \citet{Allen+08} grids computed for low metallicities using the abundances of \citet{Gutkin+16} (section \ref{sec:GutkinGrid})
  \item An extension of the \citet{Allen+08} grids computed for different shock age (section \ref{sec:particular_grids}).
\end{enumerate}

 Not all the parameter grids presented in this paper are  equally set in the shock parameter space (i.e., the size of the intervals and the range they cover may depend differently on the assumed abundances). The next section will describe the grids of models currently available in our database. For all models of the grid, we provide information about each of the three region types: shock, precursor and shock+precursor. The effect of dust have not been considered while computing the models.

\subsection{The \citet{Allen+08} grid}
\label{sec:AllenGrid}

As in \citet{Allen+08}, each shock model from the grid is uniquely defined through five different input parameters: the shock velocity \vs, the preshock transverse magnetic field $B_{0}$, the preshock density $n_{0}$ and one of the five abundance sets listed in Table\,\ref{tab:abundTableAllen}. The temperatures of the ions and electrons have been set to be equal right from the shock front. The ionization state of the precursor (the gas entering the shock front) was calculated with \MV using the method described in Sect.\,\ref{sec:preionization}.

The grid is divided into two sub-grids. Both of which are an exact replica of those presented by \citet{Allen+08} as they use the exact same shock parameters. The only exception is the use of \MV\ instead of \MIII\ during model evaluations. Below is a resum\'{e} of each sub-grid.

The first sub-grid contains 1440 models that were calculated using one of the following five abundance sets: a depleted solar set and a twice solar set which were both used by \citet{Dopita+1996}, a solar abundance set labelled 'dopita2005' from \citet{Asplund+05} that was used in \citet{Dopita+05}, and finally, an SMC and an LMC abundance set published by \citet{Russell+1992}. The respective abundances for each atomic element with respect to hydrogen are given in Table\,\ref{tab:abundTableAllen}. Each model in this sub-grid was calculated using a fixed preshock density of $n_{0} = 1$\,\cmc\ consisting of 36 individual shock velocities (from 100 up to 1000\,\kms\ in steps of 25\,\kms) and one of the following 8 transverse magnetic field values: $B$ = $10^{-4}$, 0.5, 1.0, 2.0, 3.23, 4.0, 5.0 and 10\,\muG. 

\begin{table}[ht]
\scriptsize
\caption{Abundances used in Allen et al. 2008.}
\label{tab:abundTableAllen}
\begin{center}
\begin{tabular}{lccccc}
\hline
Elem & Solar & 2$\times$Solar & Dopita2005 & LMC & SMC \\
\hline
H &  0.00 &  0.00 &  0.00 &  0.00 &  0.00 \\
He & -1.01 & -1.01 & -1.01 & -1.05 & -1.09 \\
C  & -3.44 & -3.14 & -4.11 & -3.96 & -4.24 \\
N  & -3.95 & -3.65 & -4.42 & -4.86 & -5.37 \\
O  & -3.07 & -2.77 & -3.56 & -3.65 & -3.97 \\
Ne & -3.91 & -3.61 & -3.91 & -4.39 & -4.73 \\
Na &       &       & -6.35 & -4.85 & -5.92 \\
Mg & -4.42 & -4.12 & -5.12 & -4.53 & -5.01 \\
Al & -5.43 & -5.23 & -7.31 & -4.28 & -5.60 \\
Si & -4.45 & -4.15 & -5.49 & -5.29 & -4.69 \\
S  & -4.79 & -4.49 & -5.01 & -5.23\footnotemark & -5.41 \\
Cl &       &       & -6.70 &       & -7.30 \\
Ar & -5.44 & -5.14 & -5.44 & -5.71 & -6.29 \\
Ca & -5.88 & -5.58 & -8.16 & -6.03 & -6.16 \\
Fe & -4.63 & -4.33 & -6.55 & -4.77 & -5.11 \\ 
Ni &       &       & -7.08 & -6.04 & -6.14 \\
\hline
X  & 0.7073 & 0.6946 & 0.7158 & 0.7334 & 0.7535 \\ 	
Y  & 0.2745 & 0.2696 & 0.2778 & 0.2596 & 0.2432 \\
Z  & 0.0183 & 0.0358 & 0.0065 & 0.0070 & 0.0033 \\
\hline
\hline
\end{tabular}

\end{center}
\end{table}
\footnotetext{A typo error shows -7.23 in the \citet{Allen+08} paper although their model has been obtained with the correct value of -5.23, as the one present in the grid.}

The second sub-grid of models was calculated using only the solar abundance set of \citep{Dopita+1996} with 6 preshock densities (0.01, 0.1, 1.0, 10, 100 and 1000 \cmc), at 36 shock velocities (100 up to 1000 \kms\ in steps of 25\,\kms) and for 8 transverse magnetic field values ($B$ = $10^{-4}$, 0.5, 1.0, 2.0, 3.23, 4.0, 5.0 and 10\,\muG). As in the \citet{Allen+08} paper, further models were computed using additional values of $B$ in order to cover all the magnetic parameter values $B/n^{1/2}$ from the first sub-grid. This facilitates comparison of models that differ by their preshock density. We recall that models with the same ratio $B/n^{1/2}$ result in the same magnetic to gas pressure ratio. 
These additional values (transverse magnetic field) are $B \sim 10^{-3}$, $\sim 10^{-2}$, $\sim 10^{-1}$, 1.0, 10 and 100\,\muG\ which were calculated for each selected preshock density.

\subsection{Comparison between the Allen 2008 models and our calculations}
\label{sec:compaMapps}

\begin{figure*}
  \includegraphics{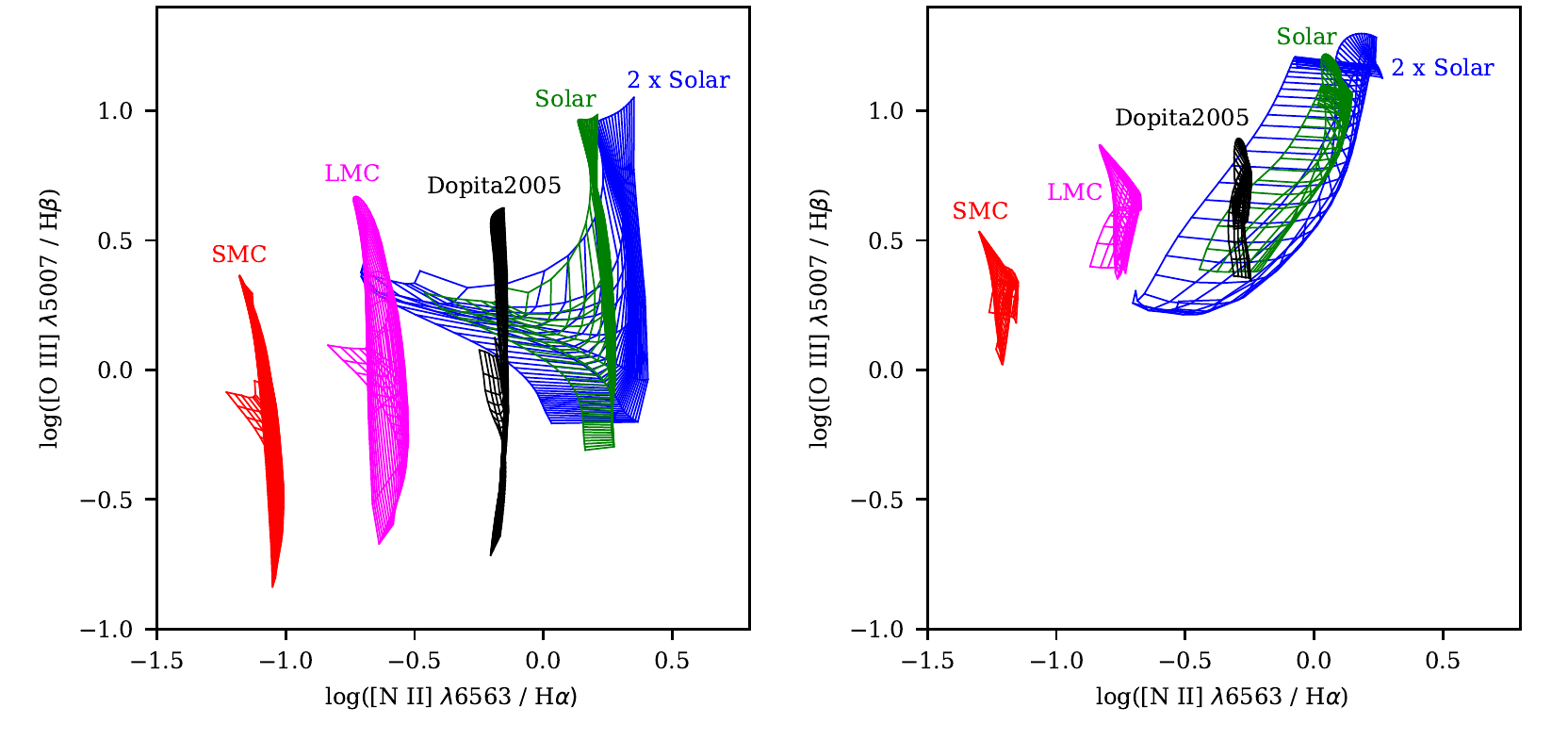}
  \caption{The BPT \OIII $\lambda$5007/H$\beta$ versus \NII $\lambda$6583/\Ha\ diagnostic diagram \citep{Baldwin+1981} displaying shock models that use the same abundance sets as \citet{Allen+08} and which cover shock velocities ranging from 200 to 1000\,\kms, all with the same preshock density of $n_0 = 1$\,\cmc. The left panel displays the line ratios from the shocked gas only while the right panel shows the same ratios after summing up shock and precursor line intensities. This figure is identical to Fig.\,20 shown in \citet{Allen+08} except that the models shown here were calculated with \MV\ instead of \MIII.}
\label{fig:BPTAllen}
\end{figure*}

The first application of the new model grid was to compare it with the previous grid from \citet{Allen+08}. Figure\,\ref{fig:BPTAllen} shows a classical plot of \OIII/\Hb\ versus \NII/\Ha, commonly known as a BPT diagnostic diagram \citep{Baldwin+1981}. Except that our grid was computed using the code \MV, it is otherwise equivalent to the one presented by \citet{Allen+08}. We assumed the same shock parameters: a solar abundance set with shock velocities varying from 200 to 1,000 \kms. The transverse magnetic field covers the range 0.0001 to 2\,\muG\ and the preshock density is always 1\,\cmc. The left panel displays the line ratios from the shocked gas only while the right panel displays the same ratios after summing up both shock and precursor line intensities. Our Figure can be directly compared to Fig. 20 of \citet{Allen+08}.

\begin{figure}
  \includegraphics{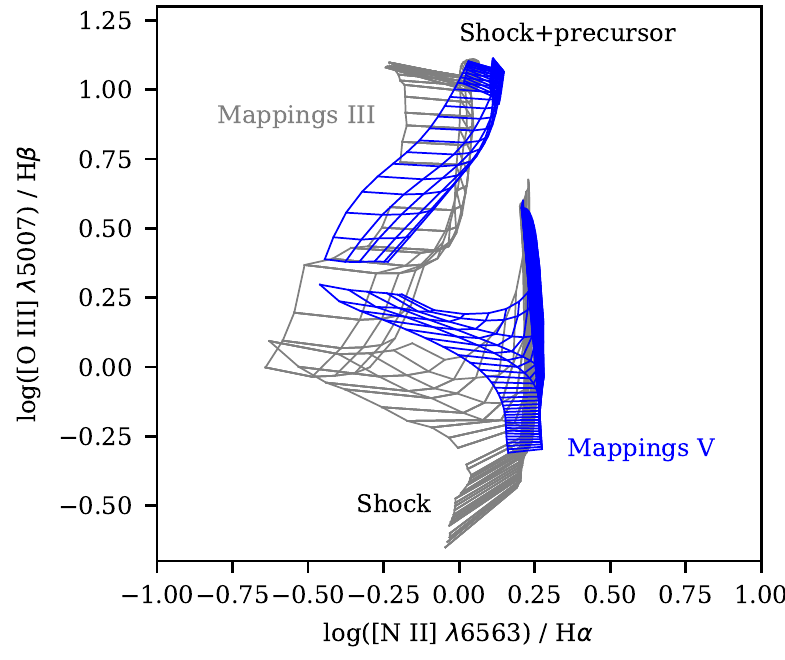}
  \caption{Comparison of \MIII\ and \MV\ in the BPT diagram of \OIII $\lambda$5007/\Hb vs. \NII $\lambda$6583/\Ha when using the same model parameters as \citet{Allen+08}: a solar abundance set, a preshock density of $n_0 = 1$\,\cmc, shock velocities ranging from 200 to 1\,000\,\kms with an interval of 25\,\kms, and a transverse magnetic field  of 10$^{-4}$, 1, 2 and 4 \muG cm$^{3/2}$. The models calculated by \citet{Allen+08} are shown in gray and our models using \MV\ are shown in blue. This figure is similar to Fig.\,21 of \citet{Allen+08}, which was used by the authors to compare their \MIII\ grid to an earlier grid from \citet{Dopita+1996}.}
 \label{fig:comparisonMappingsVersions}
\end{figure}

The Figure~\ref{fig:comparisonMappingsVersions} makes a direct comparison between our grid (blue lines) and the \citet{Allen+08} version using \MIII (gray lines). Although  similarities do appear between the different line ratio curves, a higher value by up to $\sim 0.3$\,dex of our \OIII/\Hb\ ratio shows up in some parts of the diagram.

\begin{figure*}
  \includegraphics{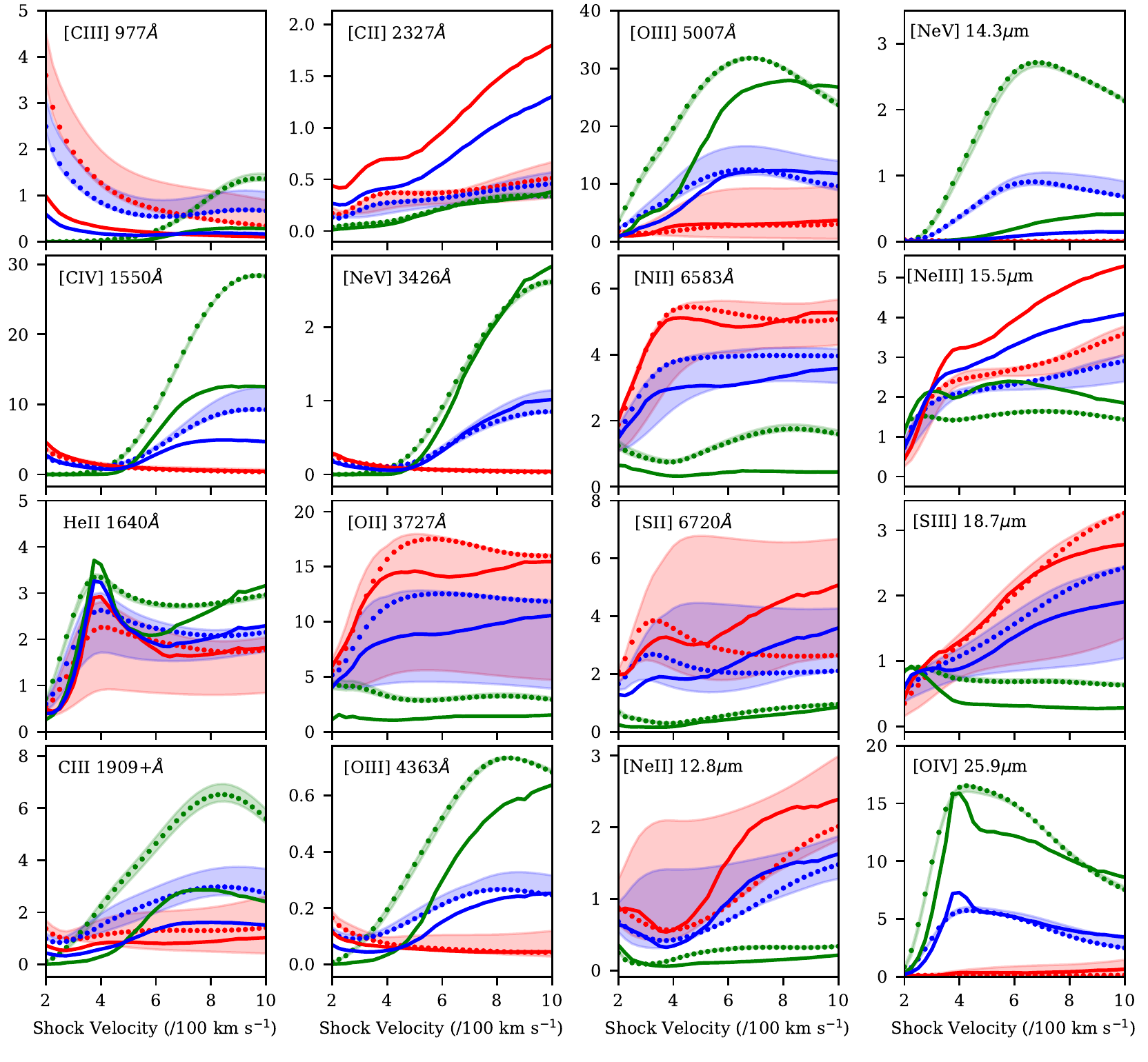}
  \caption{Line fluxes as a function of shock velocity of 16 emission lines normalized to H$\beta$. Colors represent whether the emission is from the shocked gas only (red), or from the preshock gas (green) or from the sum of both (blue). The models using \MV\ are shown using dotted lines while the continuous lines represent the \citet{Allen+08} models. The color-shaded bands represent the domain of line intensity variations of the \MV\ grid when gradually varying the magnetic field from 10$^{-4}$ to 10\,$\mu$G (with the same color coding as for the dotted lines).} 
  \label{fig:Allen08Diff}
\end{figure*}

The Figure~\ref{fig:Allen08Diff} displays the behaviour as a function of shock velocity \vs\ of 16 different emission lines. These further illustrate the similarities and differences between both grids of models. All lines are normalized to the H$\beta$ intensity. Colors represent whether the emission is from the shocked gas only (red), or from the preshock gas (green) or from the sum of  both (blue). The models using \MV\ are shown using dotted lines while the continuous lines represent the \citet{Allen+08} models. For both grids, a preshock density of 1\,cm$^{-3}$ was assumed and a magnetic field of 3.23 $\mu$G. The color shaded bands represent the range in flux variations within the \MV\ grid when the magnetic field is varied between 10$^{-4}$ and 10\,$\mu$G.  

We note from Figure~\ref{fig:Allen08Diff} that the behaviour of the emission line intensities are quite similar between the \MV\ and \MIII\ grids although significant differences appear in the case of the lines \forba{C}{iv}{1550}, \sforba{C}{iii}{1909}, \forba{O}{iii}{4363}, \forba{O}{iii}{5007}, and \forbm{Ne}{v}{14.3}, which is predicted stronger in our new grid, while the lines \forba{C}{ii}{2327} and \forbm{Ne}{iii}{15.5} are predicted weaker. Other lines appear stronger or weaker depending on the shock velocity considered.

\begin{figure*}
\centering
  \includegraphics{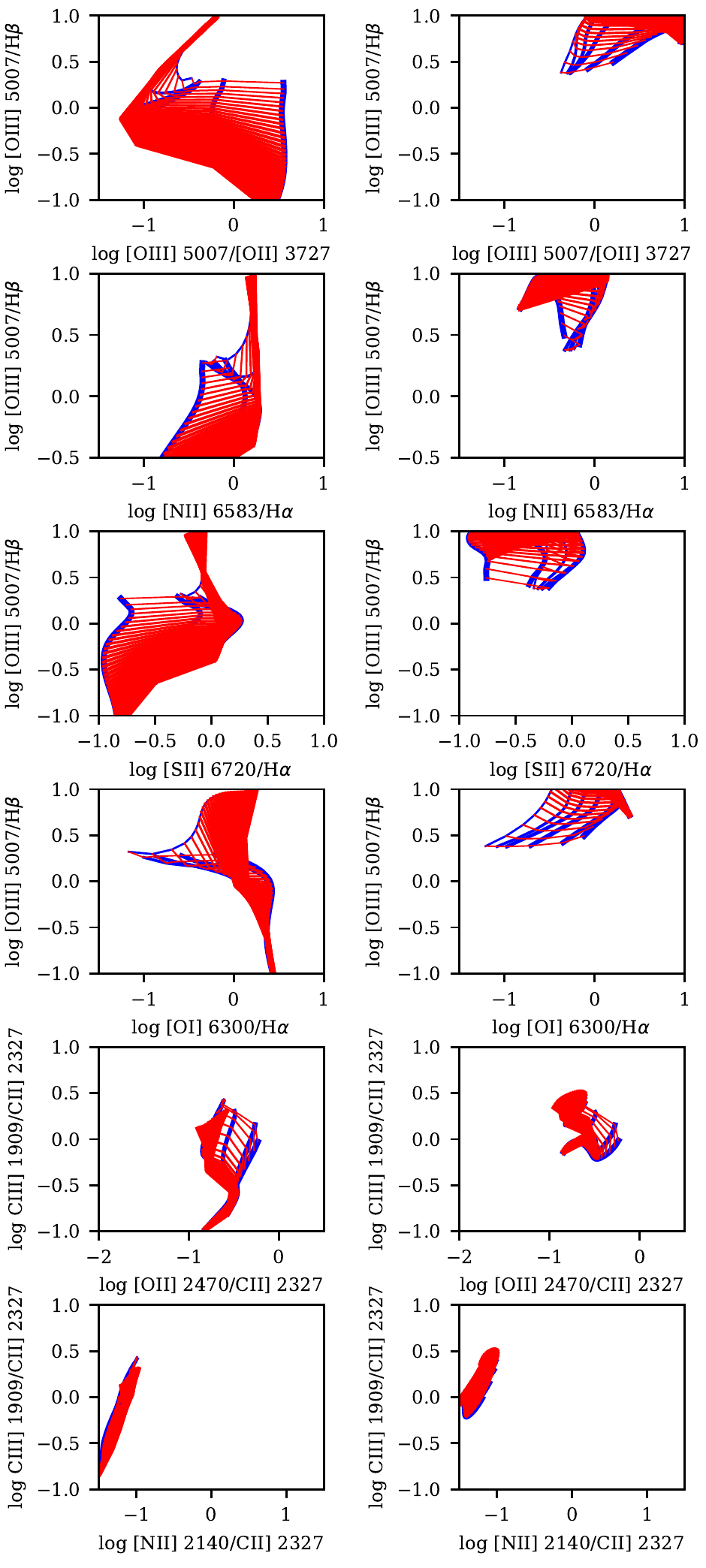}
  \caption{Diagnostic diagrams to be compared to \citet{2018ApJ...864...90M} who used \citet{Allen+08} models. All models shown have the same transverse magnetic field of 1\,$\mu$G. Red lines represent shock model sequences of different $n_0$ values but of equal velocity while blue lines represent shock model sequences different in \vs\ but of equal precursor density. The line's thickness (of the red and blue lines) increases from one sequence to the next: the blue lines getting thicker as the density gets larger and the red lines getting thicker as the velocity gets larger.}
  \label{fig:Molina}
\end{figure*}

Figure~\ref{fig:Molina} is another illustration of the differences between \citet{Allen+08} models and the current grid. The six panel pairs represent useful line ratio diagnostics when studying LINERs. All the sequences shown are from our grid only and they all have the same transversal magnetic field of 1\,$\mu$G. The emission lines from the left panels represent shocked gas only while those from the right panel represent the sum of shock with precursor emission lines. The model sequences shown correspond to sequences in which either the precursor density (red lines) or the shock velocity (blue lines) are varied. The line thickness (of the red and blue lines) increases along as the iso-parameter takes on larger values. These models are equivalent to those presented by \citet{2018ApJ...864...90M} in their figures 24 to 29. The main difference found in our grid is the \forba{O}{iii}{5007} line, which can be up to 0.3\,dex stronger than \citet{2018ApJ...864...90M}.

The line ratio differences discussed above between code versions arise either from changes implemented in the newer \M\ code  or from the use of a more recent atomic database. It is beyond the scope of this paper (and of the authors expertise) to determine which of these is at play in any specific line intensity difference.

\subsection{The low metallicity grid}
\label{sec:GutkinGrid}

We have extended the grid of \citet{Allen+08} to include low metallicity abundances in order to study shocks in galaxies at different cosmic epochs. The abundance sets chosen for this grid in particular are the same as the ones derived in \citet{Gutkin+16}, which were evaluated for different mass fraction of Z. We follow the same methodology, which these authors described in their paper, in order to retrieve the abundance of each individual element, from hydrogen to zinc. 

\begin{figure*}[ht]
  \includegraphics{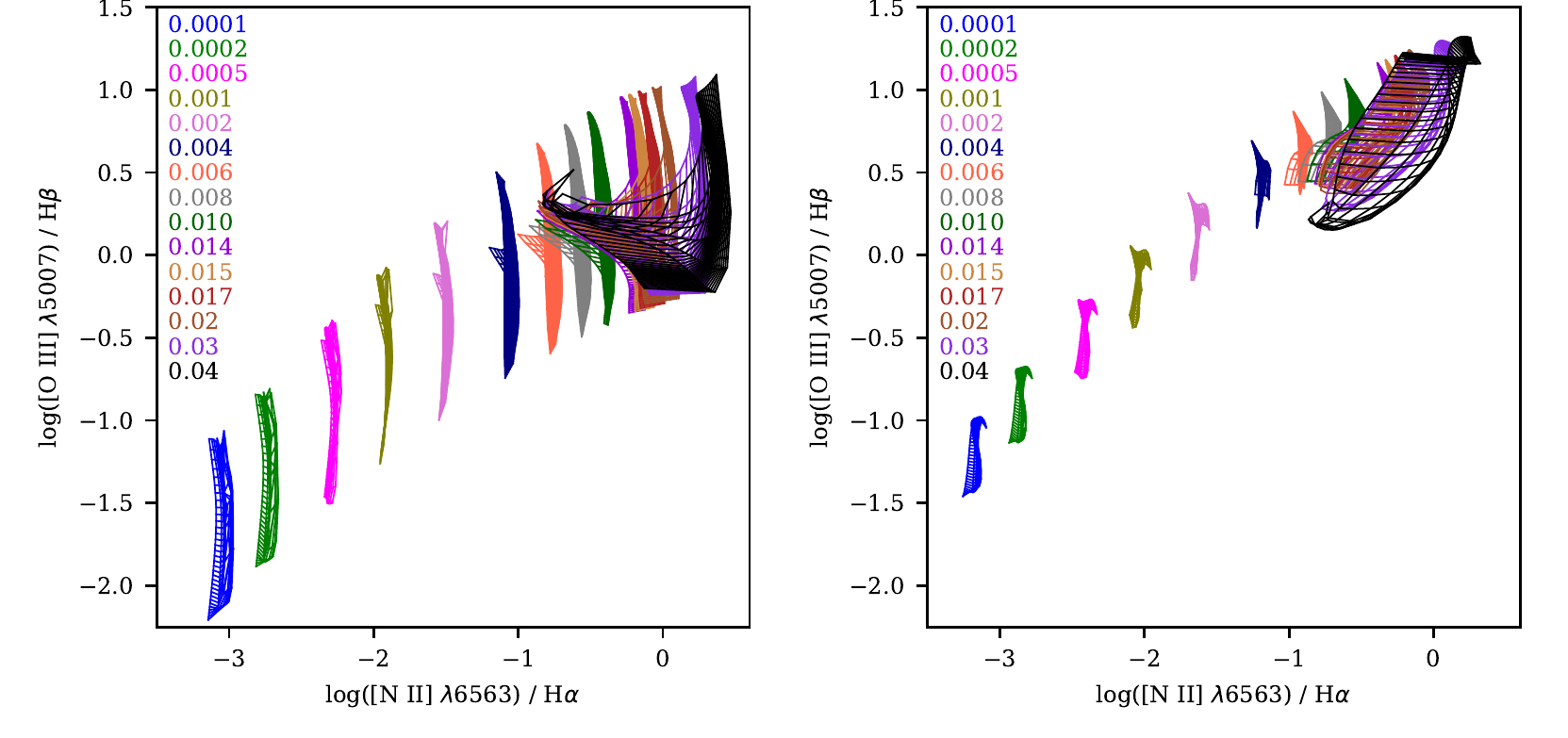}
  \caption{Same diagrams as in Fig.\ref{fig:BPTAllen} except that the abundance sets are now from \citet{Gutkin+16}. Shock velocities range from 100 to 1\,000\,km/s, and magnetic fields range from 0.0001 to 10 \muG. The left panel corresponds to the line emission ratios from the shocked gas while the right panel represents the line emission of both the shocked and precursor gas. The column in the upper left corner of each panel lists the mass fractions of each sequence of the corresponding color.}
  \label{fig:GutkinBPT}
\end{figure*}

Our database currently contains models calculated using the parameters given in Table\,\ref{tab:gridGutkin}. Figure~\ref{fig:GutkinBPT} allows us to visualize the effect on the BPT diagram \OIII/\Hb vs. \NII/\Ha  of adopting lower values of Z, starting with a value of (C/O)/(C/O)$_{\odot}$ = 1 and successively going down to $10^{-4}$ (shifting from left to right in the figure). Interestingly, when the metallicity decreases, the horizontal spread of \NII/\Ha\ in each sequence tends to become narrower. Both changes of the shock velocity (from 100 to 1\,000\,km/s) and the magnetic field (from 10$^{-4}$) to 10 \muG) lead to an increase in \forba{O}{iii}{5007}/H$\beta$, while \forba{N}{ii}{6583}/H$\alpha$ remains essentially constant.
It is also apparent from these figures that very low metallicity shocks lead to very low values of any collisionally excited line when normalized with respect to a recombination line of H. It is important to keep in mind that at the very low metallicity end, the lines from metals become negligible, but the hydrogen and helium recombination or collisionally excited lines are still emitted. If such shocked gas emission was superposed to the observation of a photoionized region, it might lead to an underestimation of metallic abundances that would unavoidably result from simply assuming classical methods to determine e.g. O/H.

\begin{table}
\scriptsize
\caption{Grid sample of the low metallicity grid described in section 3.2}
\label{tab:gridGutkin}
\begin{center}
\begin{tabular}{ll}
\hline
Parameter & Sampled values \\
\hline
(C/O)/(C/O)$_{\odot}$ &  0.26, 1.00 \\
Z$_{ism}$ &  0.0001, 0.0002 0.0005, 0.001, 0.002, 0.004 \\
 		  & 0.006, 0.008, 0.010, 0.014, 0.01524, 0.017 \\
 		  & 0.02, 0.03, 0.04 \\
v$_{s}$ (km s$^{-1}$) & 100, 125, ..., 1000 \\
n$_{0}$ (cm$^{-3}$) & 1, 10, 10$^{2}$, 10$^{3}$, 10$^{4}$ \\
B$_{0}$ ($\mu$G) & 10$^{-4}$, 0.5, 1.0, 2.0, 3.23, 4.0, 5.0, 10  \\
\hline
\end{tabular}
\end{center}
\end{table}

\subsection{Grids of truncated/young shock models}
\label{sec:particular_grids}

The \citet{Allen+08} fast shock grid (sec.\,\ref{sec:AllenGrid}) and the low metallicity shock grid (sec. \ref{sec:GutkinGrid}) contain  models for which both the age of the shock was fixed prior to the calculations by assuming arbitrary predetermined values. In certain cases, this parameters, when treated as free quantity turn out to influence the line intensities in ways that can be interesting to explore. In this section, we will indicate how the database can be used to explore the effect of the shock age on line ratios. 

The great majority of shock models found in the database are complete models, that is, they have been fully integrated until the shocked gas has cooled and fully recombined. They correspond to steady-state conditions and such shocks with time will slow down as they progressively convert their supersonic kinetic energy into radiative cooling. Although steady-state models can reproduce the conditions encountered in a wide variety of astrophysical objects, they are not always the optimal perspective. This is the case for instance in objects in which the shocks are relatively recent and therefore possess an incomplete cooling structure. For this particular case, a grid of incomplete cooling shock models, also known as truncated models or young shocks, is needed. Only a few calculations of young shocks can be found in the literature. They were computed to match individual object such as the Cygnus Loop filaments \citep{Raymond+1980, Contini+1982, Raymond+1988} or low-excitation Herbig-Haro objects \citep{Binette+1985}. The lack of grids that included young/truncated shocks has compelled us to add those to our database and to provide a way of exploring them.

\begin{figure*}[ht]
  \includegraphics{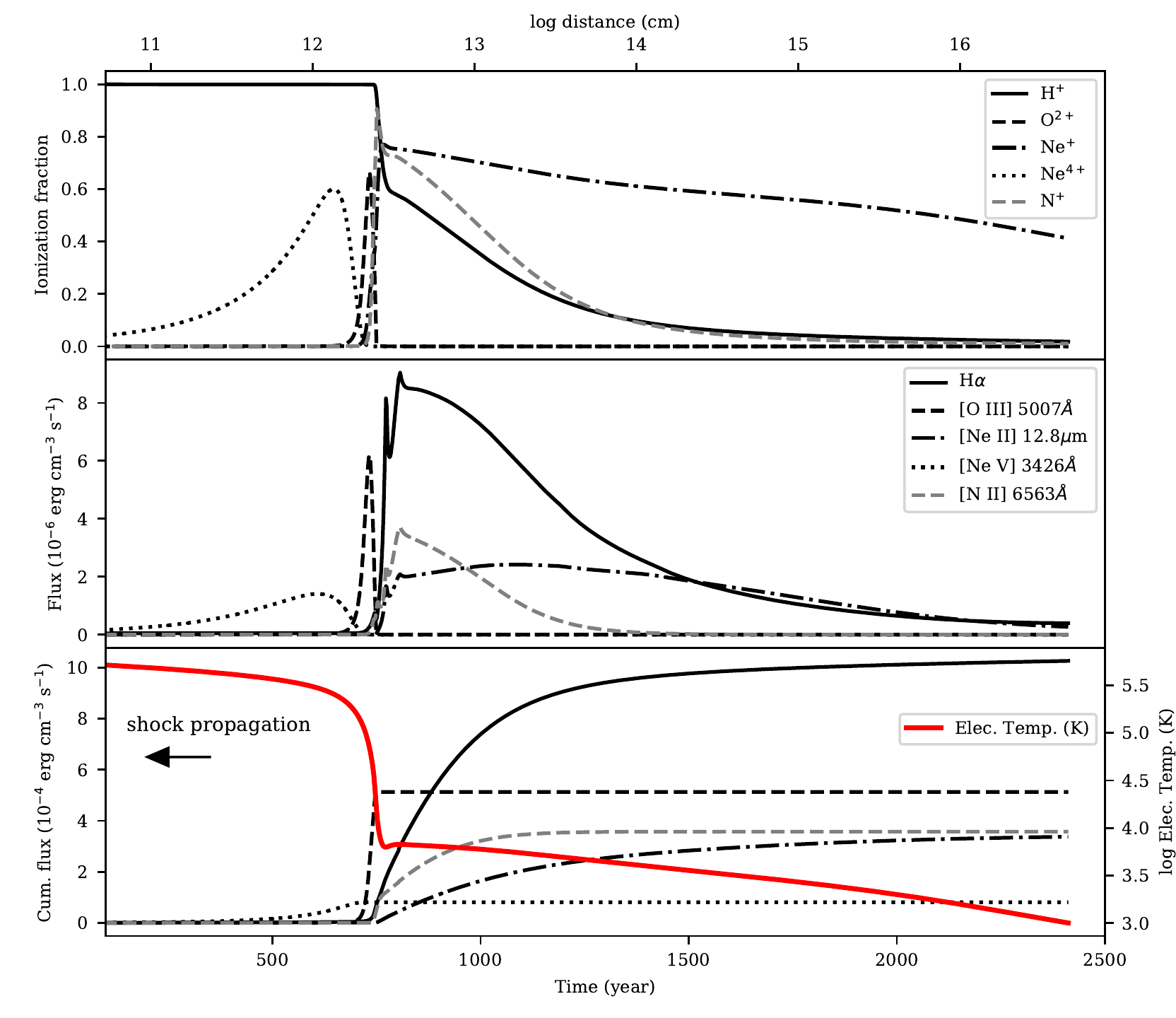}
  \caption{Example of an ionization and emissivity structure for different element at different ionization stages behind a 200\,\kms shock propagating into a preshock density of 10\,\cmc and a transverse magnetic field of 1\,\muG with solar abundances. Top panel: Ionization fraction of H\,II, O\,III, Ne\,II and Ne\,V. Middle: emissivity of H$\alpha$, \forba{O}{iii}{5007}, \forbm{Ne}{ii}{12.8}, and \forba{Ne}{v}{3426}. Bottom panel: cumulative emissivity for the same ions as plotted in the middle panel. The shock front is propagating towards the left.}
  \label{fig:ionization_profile}
\end{figure*}

\begin{figure*}[ht]
  \includegraphics{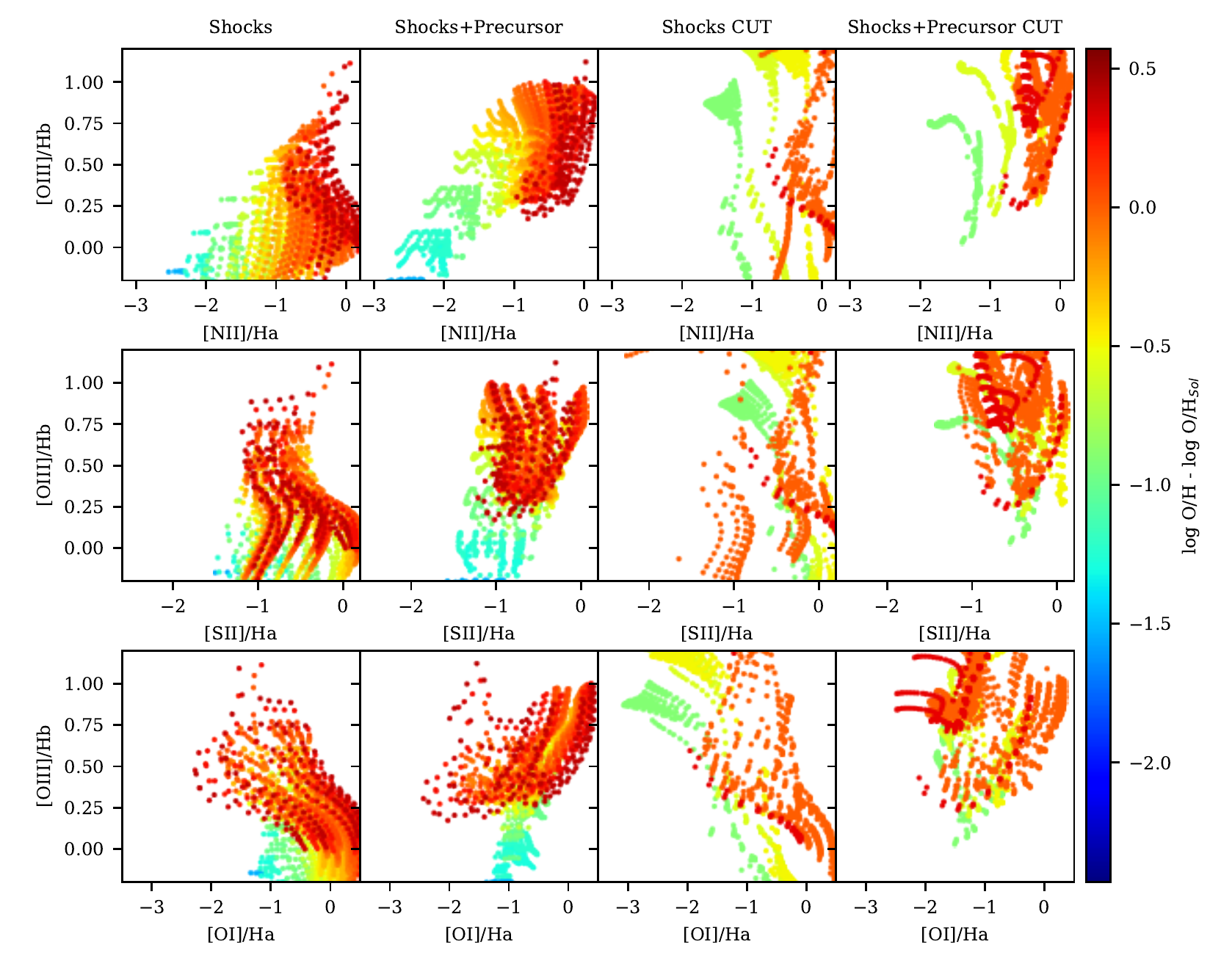}
  \caption{Three classical BPT-type diagrams that compare complete with incomplete shock models (labeled "CUT"). All models were evaluated using a transversal magnetic field of 1 \muG. The panels depict line ratios from either the shocked gas only or from both the shocked gas with the precursor emission.}
  \label{fig:cutShocks}
\end{figure*}

Fig.~\ref{fig:cutShocks} illustrates the impact on the line ratios of using incomplete shocks. It can be compared to the similar Fig.\,6 of \citet{Kehrig+2018}. A very common assumption is that when one observes high values of \SII or \NII or \OII lines when normalized by an H line, this implies the presence of shocks or, on the contrary, when these ratios turned out small, one often concludes that shocks cannot be involved in the gas excitation. Such generalized stands must be applied with greater care. For instance, at low metallicities, we find that young shock models can easily fall under the classical Kewley curve \citep[see][]{Kewley+2001}.

Truncated (or age-limited or incomplete) shock models can be identified in the database using the \sqlTab{cut} keyword as a suffix at the end of the grid reference (ex: Allen08-cut, the only ones available for the moment). Before using any such model, it is important to be fully aware of what an incomplete or age-limited shock represents, which implies becoming familiar with the behaviour of the ionization, temperature and line emissivities downstream from the shock front. To give an idea, Fig. \ref{fig:ionization_profile} shows how the evolution of these quantities are structured for a 200\,\kms shock propagating into a gas with solar abundances and preshock density of 10\,\cmc. As it can be seen on this figure, the recombination occurs in stages, with the higher ionization species recombining towards lower values as time proceeds. The gas temperature shown in the bottom panel declines markedly with time. Both of these factors and the fact that the density increases as the temperature drops\footnote{The transverse magnetic field will provide pressure to the gas as the temperature drops and if intense enough will suspend the usual isobaric prescription.} means that the emissivity of any given line strongly varies with time along the cooling history of the postshock gas. The lower panel in Fig.\,\ref{fig:ionization_profile} shows how the cumulative (time-integrated) flux emission progresses, up to the desired final shock age. These time-integrated emission fluxes are actually stored in the database as a function of time. As revealed by the bottom panel, the line intensities are very sensitive to the time spent since the passage of the shock front. It is highly recommended to explore how such variations in line intensities come about and from there select the appropriate parameters to vary. 

There are alternative parameters that can be used to specify the degree of shock completeness. The first is obviously the time elapsed (age) since the shock front initiated. The age parameter is located in the \sqlTab{shock\_params} table in column named \sqlTab{time}, which is expressed in seconds. There is also the distance between the shock front and the final thickness of the shocked gas (at the specified shock age). This parameter is similarly located in the \sqlTab{shock\_params} table in the column named \sqlTab{distance}, which is expressed in cm. Finally, there is the integrated column density of each ionic specie  of the gas swept by the shock front until the gas has cooled and recombined (column named \sqlTab{coldens}), which is expressed in cm$^{-2}$. The total H column density can be found with the following summation : $N_H = N_{HI} + N_{HII}$. Complementary evaluations of how a specific line emissivity varies with age or depth requires knowing the integrated column density of the ion involved, which can be found in the \sqlTab{ion\_col\_dens} table. 

\begin{figure}
  \includegraphics{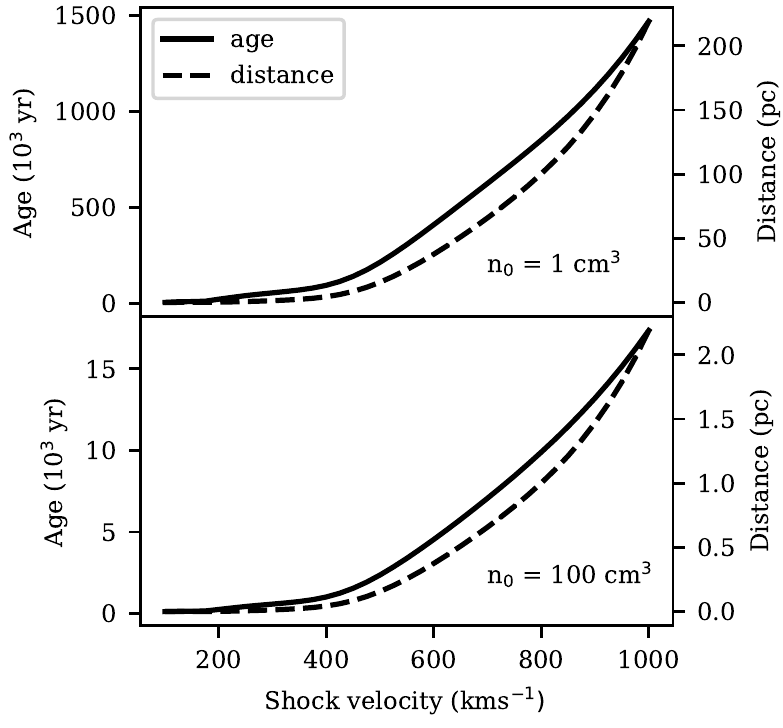}
  \caption{Age and thickness of two shocks of different velocities that propagate in a gas cloud of density of either 1\,\cmc (top panel) or  100\,\cmc gas (bottom panel), assuming in both cases solar abundances and B$_{0}$ = 1\,\muG.}
  \label{fig:distance_age_shock}
\end{figure}

Both the distance travelled by the shock wave or alternatively its cumulative age can vary greatly depending on the initial shock conditions has shown in Fig. \ref{fig:distance_age_shock}. An easy mistake a user can make is to define diagnostic diagrams such as those presented in Figs.\,\ref{fig:BPTAllen} and \ref{fig:GutkinBPT} and superpose them to observations characterized by ages and/or sizes that are inconsistent with those of the models. 
This can be avoided by varying the preshock density until its thickness fits the desired scale. In the case of a complete shock model (without an external photoionization source), the final thickness and the time spent in cooling both scale as $(n_0)^{-1}$ (as seen in Fig.\ref{fig:distance_age_shock}).

Another potential mistake would be to use too high a preshock density, resulting in models where the density one would infer from its \SII ($\lambda$6716\AA/$\lambda$6731\AA) doublet ratio (or its \OII ($\lambda$3726\AA/$\lambda$3729\AA) doublet ratio) are much higher than the density values derived from the observations. Consider for instance a postshock temperature of $10^{5.3}$\,K without magnetic field will result in a \SII doublet ratio corresponding to a density as high as $\sim$ 100 $n_0$. The reason is that the density increases isobarically as the shock cools and, for the $B=0$ case, the local density across the whole shock structure grows as the ratio $T_{post}/T_e$ where $T_{post}$ and $T_e$ are the postshock and the local electronic temperature, respectively.

\section{Future evolution of the database}
New models will be added depending on the research necessities of the authors of this paper or upon request by anybody interested in models that have not yet been included. As the web page is dynamic and  will instantly reflect the current status of the database and of its model grids, any new references or model inclusions will be easily noticed. The authors of this paper will also do their best to update the 3MdBs grid of models as soon as possible after a new version of \M is available. 

\section{Concluding remarks}

In this paper, we have presented an extension to the Mexican Million models database (3MdB) \citep{Morisset+2015} which comprises of the addition of shock models to the database. Important information were presented in this paper and the following is a list that summarized the important points :

\begin{enumerate}
 	 \item All shock models in the database were computed using the shock modelling code \MV. 
 	 
 	 \item The database structure has been explain in detail and information about how to connect and interact with the database has been covered. A website was created in order to follow the evolution of the database as new grids will be added in the future.
 	 
 	 \item At the time of publication, 3 grids were available. 1 : The grid of \citet{Allen+08} was re-computed using \MV. 2 : We have extended the grids of \citet{Allen+08} to low metallicity using the abundances derived by \citet{Gutkin+16}. 3 : Grids of truncated/young shock models were computed. 
\end{enumerate}

\section{Acknowledgments}
The current work is dedicated to the memory of Mike Dopita, the creator of \textsc{mappings}. The authors wish to thank Luc Binette for his very useful collaboration during this study and the redaction of this paper.
The current work was presented at the conference ${A\,star\,was\,born}$ celebrating the scientific achievements of Mike Dopita in April 2018. The 3MdBs access tools have been developed by A. Alarie while he was funded by a postdoctoral grant from CONACyT.
This work is supported by grants DGAPA/PAPIIT-107215 and CONACyT-CB2015-254132. 

\bibliography{main}

\appendix

\begin{table*}[ht]
\caption{Field list of Table "shock\_params"}
\label{tab:shockParamsTable}
\begin{center}
\begin{tabular}{cl}
\hline
Field name & Description \\
\hline
\sqlTab{ModelID} & Unique model identification number$^{*}$\\
\sqlTab{ProjectID} & Unique project identification number$^{*}$ \\ 
\sqlTab{AbundID} & Abundance identification number (same as in abundances table) \\
\sqlTab{FHI} & Ionization fraction of H$^{0}$ \\
\sqlTab{FHII} & Ionization fraction of H$^{+}$ \\
\sqlTab{FHeI} & Ionization fraction of He$^{0}$ \\
\sqlTab{FHeII} & Ionization fraction of He$^{+}$ \\
\sqlTab{FHeIII} & Ionization fraction of He$^{++}$ \\
\sqlTab{shck\_vel} & Shock velocity (\kms) \\ 
\sqlTab{preshck\_dens} & Preshock density (\cmc) \\ 
\sqlTab{preshck\_temp} & Preshock electronic temperature (K) \\
\sqlTab{mag\_fld} & Transverse magnetic field (\muG) \\
\sqlTab{cut\_off\_temp} & Final electronic temperature in the last zone evaluated (K) \\
\sqlTab{ref} & Name of the grid in which the model is belonging \\
\sqlTab{script} & Main script used to evaluate a model$^{*}$ \\
\hline
\end{tabular}
\begin{tablenotes}
  \small
  \item $^{*}$ Field used internally in the web application. 
   \end{tablenotes}
\end{center}
\end{table*}
\begin{table*}[ht]
\caption{Field list of Table "model\_directory"}
\label{tab:model_directory}
\begin{center}
\begin{tabular}{ll}
\hline
Field name & Description \\
\hline
\sqlTab{Created} & Date the model was added to the database \\
\sqlTab{ModelID} & Unique model identification number \\
\sqlTab{Parameters} & Model parameters associated with the grid (shock or photoionization$^{*}$) \\
\sqlTab{ProjectID} & Unique project identification number \\
\sqlTab{code\_version} & \textsc{mappings} version used \\
\hline
\end{tabular}
\begin{tablenotes}
  \small
  \item $^{*}$ To be implemented in the future.
   \end{tablenotes}
\end{center}
\end{table*}
\begin{table*}[ht]
\caption{Field list of Table "projects"}
\label{tab:projectTable}
\begin{center}
\begin{tabular}{ll}
\hline
Field name & Description \\
\hline
\sqlTab{Created} & Date the grid was added to the database \\
\sqlTab{ProjectID} & Unique project identification number \\
\sqlTab{code\_version} & \textsc{mappings} version used \\
\sqlTab{model\_count} & number of models for this project in the database \\
\sqlTab{ref} & Name of the grid \\
\hline
\end{tabular}
\end{center}
\end{table*}
\begin{table*}[ht]
\caption{Field list of Table "abundances"}
\label{tab:abundancesTableFields}
\begin{center}
\begin{tabular}{ll}
\hline
Field name & Description \\
\hline
\sqlTab{name} & Abundances file name used during the evaluation\\
\sqlTab{AbundID} & Abundance identification number \\
\sqlTab{X} & Hydrogen mass fraction \\
\sqlTab{Y} & Helium mass fraction \\
\sqlTab{Z} & Metallicity mass fraction of all element heavier than helium \\
\sqlTab{HELIUM} & Abundance of helium in log(He/H) \\
\sqlTab{LITHIUM} & Abundance of lithium in log(Li/H) \\
\sqlTab{BERYLLIUM} & Abundance of beryllium in log(Be/H) \\
\sqlTab{BORON} & Abundance of boron in log(B/H) \\
\sqlTab{CARBON} & Abundance of carbon in log(C/H) \\
\sqlTab{NITROGEN} & Abundance of nitrogen in log(N/H) \\ 
\sqlTab{ZINC} & Abundance of helium in log(Zn/H) \\
\hline
\end{tabular}
\end{center}
\end{table*}
\begin{table*}[ht]
\caption{Field list of Table "ion\_frac", "ion\_temp" and "ion\_col\_dens"}
\label{tab:ionFractionTable}
\begin{center}
\begin{tabular}{lc}
\hline
Field name & Ionization range$^{*}$ \\
\hline
\sqlTab{HYDROGEN} & (0 to 1) \\
\sqlTab{HELIUM} & (0 to 2) \\
\sqlTab{BERYLLIUM} & (0 to 4) \\
\sqlTab{BORON} & (0 to 5) \\
\sqlTab{CARBON} & (0 to 6) \\
\sqlTab{NITROGEN} & (0 to 7) \\
\sqlTab{OXYGEN} & (0 to 8) \\
\sqlTab{FLUORINE} & (0 to 8) \\
\sqlTab{NEON} & (0 to 8) \\
\sqlTab{SODIUM} & (0 to 9) \\
\sqlTab{MAGNESIUM} & (0 to 9) \\
\sqlTab{ALUMINIUM} & (0 to 9) \\
\sqlTab{SILICON} & (0 to 9) \\
\sqlTab{PHOSPHORUS} & (0 to 9) \\
\sqlTab{SULPHUR} & (0 to 9) \\
\sqlTab{CHLORINE} & (0 to 9) \\
\sqlTab{ARGON} & (0 to 9) \\
\sqlTab{POTASSIUM} & (0 to 10) \\
\sqlTab{CALCIUM} & (0 to 10) \\
\sqlTab{SCANDIUM} & (0 to 12) \\
\sqlTab{TITANIUM} & (0 to 12) \\
\sqlTab{VANADIUM} & (0 to 13) \\
\sqlTab{CHROMIUM} & (0 to 13) \\
\sqlTab{MANGANESE} & (0 to 13) \\
\sqlTab{IRON} & (0 to 13) \\
\sqlTab{COBALT} & (0 to 13) \\
\sqlTab{NICKEL} & (0 to 13) \\
\sqlTab{COPPER} & (0 to 13) \\
\sqlTab{ZINC} & (0 to 13) \\
\hline
\end{tabular}
\begin{tablenotes}
  \small
  \item $^{*}$ 0=X$^{0}$, 1=X$^{+}$, 2=X$^{++}$...
   \end{tablenotes}
\end{center}
\end{table*}

\end{document}